\RequirePackage[T1]{fontenc}
\documentclass[letterpaper, 10 pt, conference]{ieeeconf}  

\IEEEoverridecommandlockouts                              

\usepackage{graphicx} 
\usepackage{pgfplots}
\pgfplotsset{compat=newest,
			    try min ticks=3}
\usepackage{tikz}
\newlength\figh			
\newlength\figw			

\usepackage{epsfig} 
\usepackage{mathptmx} 
\usepackage{times} 
\usepackage{amsmath} 
\usepackage{amssymb}  
\usepackage{xcolor} 

\usepackage{arydshln} 
\usepackage{booktabs}
\usepackage{bm}
\usepackage[style=ieee]{biblatex} 
\title{\LARGE \bf
Transformation-Free Fixed-Structure Model Reduction for LPV Systems 
}

\author{Lennart Heeren$^{1}$, Adwait Datar$^{1}$, Antonio Mendez Gonzalez$^{1}$ and Herbert Werner$^{1}$
\thanks{$^*$ This work was partially funded by the German Research Fundation (DFG) - project number 413984989.}
\thanks{$^{1}$ Lennart Heeren, Adwait Datar, Antonio Mendez Gonzalez and Herbert Werner are with the Institute of Control Systems,
               Hamburg University of Technology, 21073 Hamburg, Germany 
               (e-mail: lennart.heeren@tuhh.de, h.werner@tuhh.de).}%
}

\begin{document}

\maketitle
\thispagestyle{empty}
\pagestyle{empty}

\begin{abstract}
In this paper, we propose a model reduction technique for linear parameter varying (LPV) systems based on available tools for fixed-structure controller synthesis.
We start by transforming a model reduction problem into an equivalent controller synthesis problem by defining an appropriate generalized plant.
The controller synthesis problem is then solved by using gradient-based tools available in the literature.
Owing to the flexibility of the gradient-based synthesis tools, we are able to impose a desired structure on the obtained reduced model.
Additionally, we obtain a bound on the approximation error as a direct output of the optimization problem.
The proposed methods are applied on a benchmark mechanical system of interconnected masses, springs and dampers.
To evaluate the effect of the proposed model-reduction approach on controller design, LPV controllers designed using the reduced models (with and without an imposed structure) are compared in closed-loop with the original model.
\end{abstract}
\section{Introduction}
Linear parameter varying (LPV) systems are a generalization of linear time-invariant (LTI) systems where the model matrices have a continuous-dependence on a so-called scheduling variable (see \cite{shamma2012overview}, \textcolor{black}{\cite{hoffmann2014survey}}).
The LPV framework offers analysis and synthesis tools to systematically design controllers with stability and performance guarantees.
However, in some cases the available techniques fail to properly
scale with the high complexity of the considered LPV system and can be intractable to apply, specially for very high order systems.

Although the literature on model reduction techniques for LTI systems is vast, it is relatively sparse for LPV systems and thus forms an active area of research.
A typical model reduction approach involves an interpolation step wrapping around standard LTI model reduction techniques applied on a grid of of scheduling parameters values. (see \cite{theis2015modal}, \cite{amsallem2011online}).
A grid-free approach to LPV model reduction is presented in \cite{heeren2022grid}.
Standard model reduction techniques based on balanced truncation involve solving a set of Linear matrix inequalities (LMIs) to obtain the generalised  controllability and observability Gramians \cite{Wood.1995}.
From the generalised Gramians a balanced realisation can be constructed, a truncation of which then yields 
 a reduced state-space  model.
The approximation error is obtained in the form of a sum of the truncated Hankel singular values.
Heuristic approaches to minimize this sum of singular values are deployed to obtain a reduced model with the minimum approximation error.
In this paper, we bypass this two-step procedure and propose to directly minimize the approximation error bound using fixed-structure controller synthesis techniques.
Following up on ideas from \cite{beck1997model} and \cite{al2016structure}, we cast the model reduction problem in the form of a fixed-structure synthesis problem by appropriately defining a generalized plant.
This opens the possibility to apply efficient tools developed for addressing the fixed-structure controller synthesis problem \cite{chughtai2011hybrid,chughtai2010simply,chughtai2007fixed,abbas2008hybrid}.
The optimization problem yields the locally optimal cost which then  provides theoretical guarantees in terms of an upper bound on the approximation error.  
Owing to the flexibility of these gradient-based synthesis tools, we are able to impose a desired structure on the obtained reduced model.
For example, it might be desirable to impose a block-diagonal modal structure on the state-matrix with each block of size $2$ such that each block defines a particular mode  of a system corresponding to a pair of complex-conjugate eigenvalues.
Such a block-diagonal structure in the state-matrix has a good physical interpretation and can be used to design different controllers for different frequency regimes.
For example, when looking at applications from structural mechanics, one is often interested in modeling and controlling a specific vibration-mode in a system.
Another application comes from aeroservoelastic vehicles where a modal structure is desirable for control design (see \cite{theis2015modal}).
This might further open doors for developing model reduction techniques that preserve the typical Lagrangian structure of mechanical systems \cite{lall2003structure}.

Thus, the problem addressed in this paper is to obtain a reduced model with guaranteed error bound when a particular model structure is imposed.

\subsection{Outline}
The outline of this paper is as follows.
We start by discussing the preliminaries in Section \ref{se:preliminaries} followed by a description of the proposed model reduction procedure in Section \ref{sec:mode_reduction_tech}.
We next apply the proposed techniques on a benchmark mass-spring-damper example in Section \ref{se:numericalResults} and finally conclude the paper in Section \ref{se:conclusion}.
\section{Preliminaries} \label{se:preliminaries}
The linear dynamics of linear parameter varying (LPV) systems depend on a vector of time-varying scheduling parameters
\begin{equation*}
   \rho(t) = [\rho_1(t),...,\rho_{n_\rho}(t)]^\text{T}.
\end{equation*}
The values of $\rho(t)$ can be measured online but are not known {\em a priori}.
The general form of an LPV system is
\begin{subequations} \label{eq:system}
   \begin{align}
         \dot{x}(t) &= A(\rho(t)) x(t) + B(\rho(t)) u(t),\\
         y(t) &= C(\rho(t)) x(t) + D(\rho(t)) u(t)
   \end{align}
\end{subequations}
with  state, input and output denoted by $x(t) \in \mathbb{R}^{n_x}$, $u(t) \in \mathbb{R}^{n_u}$ and $y(t) \in \mathbb{R}^{n_y}$, respectively.
Here $A(\rho)$, $B(\rho)$, $C(\rho)$ and $D(\rho)$ are matrix-valued continuous functions of $\rho$ and we use the notation $G=\left[\begin{array}{c|c}
	A(\rho)     &  B(\rho)\\
	\hline
	C(\rho)     &  D(\rho)
\end{array}\right]$ to denote the LPV system. 
For the method proposed in this paper, a rational dependence on $\rho$ is assumed that enables the construction of an LPV model in a Linear Fractional Transformation (LFT) form reviewed in the next sub-section.
If the dependence is not rational initially, it can be made so by introducing new scheduling parameters (possibly at the expense of some conservatism).
All admissible trajectories of the scheduling parameters are assumed to be contained in a compact set,
\begin{equation*}
   \rho(t) \in \mathcal{P} \subset \mathbb{R}^{n_\rho}, \hspace{.5cm} \forall\,t \geq 0.\rule[-1cm]{0pt}{0cm}
\end{equation*}
The goal of LPV model order reduction is to find a representation
\begin{subequations} \label{eq:reducedSystem}
   \begin{alignat}{2}
      \dot{x}_\text{red}(t) &= A_\text{red}(\rho(t)) x_\text{red}(t) \,+\, & B_\text{red}(\rho(t)) u(t),&\\
      y(t) &= C_\text{red}(\rho(t)) x_\text{red}(t) \,+\, & D(\rho(t)) u(t),&
   \end{alignat}
\end{subequations}
that approximates system~(\ref{eq:system}) in its input-output behavior while the state vector $x_\text{red}(t) \in \mathbb{R}^{n}$ is of a significantly smaller dimension ($n \ll n_x$).
To simplify notation, subscripts are used to indicate parameter dependencies, e.g. $A_\rho := A(\rho(t))$.
For a linear parameter varying system $G$, we denote by $||G||$, the induced $\mathcal{L}_2$ norm of the system which reduces to the $\mathcal{H}_{\infty}$ norm if the system is linear. 
\subsection{Linear Fractional Transformation}
\textcolor{black}{For matrices $M$, $\Delta$ and $K$ of compatible dimensions, an upper LFT denoted by $\mathcal{F}_u(M,\Delta)$ and a lower LFT denoted by $\mathcal{F}_l(M,K)$ is defined as
\begin{align*}
   \mathcal{F}_u\left(\begin{bmatrix} 
      M_{11} & M_{12}\\
      M_{21} & M_{22}
   \end{bmatrix},\Delta\right)
   &= M_{22} + M_{21} \Delta \left( I-M_{11} \Delta \right)^{-1}M_{12},\\
   \mathcal{F}_l\left(\begin{bmatrix} 
      M_{11} & M_{12}\\
      M_{21} & M_{22}
   \end{bmatrix},K\right)
   &= M_{11} + M_{12} K \left( I-M_{22} K \right)^{-1}M_{21}.
\end{align*}
Assuming that the dependence of the model matrices in (\ref{eq:system}) on $\rho$ is rational, the system can be described in LFT form
\begin{align} \label{eq:lft}
   \begin{bmatrix}
      A_\rho & B_\rho \\
      C_\rho & C_\rho
   \end{bmatrix}
   = \mathcal{F}_u\left(\left[
      \begin{array}{@{}c:cc@{}}
         D_{vw}   & C_v & D_{vu} \\
         \hdashline
         B_w      & A & B_u \\
         D_{yw}   & C_{y} & D_{yu}
      \end{array}
   \right],\Delta\right),
\end{align}}
where the diagonal matrix $\Delta(t) \subset \mathbb{R}^{(q \times q)}$ is constructed from the entries of $\rho(t)$ by using standard tools for the construction of linear fractional representations.
The compact set ${\cal P}$ is then mapped into a compact set $\bm{\Delta}$.
This representation is \textit{well-posed} if the matrix ($I-D_{vw}\Delta$) has full rank for all $\Delta \in \bm{\Delta}$ \cite{Zhou.1996}.
Note that the size $q$ of $\Delta$ can exceed the number of scheduling parameters.
The matrices $B_w$, $C_v$ and $D_{vw}$ are chosen such that $\|\Delta\| \leq 1$ for all $\Delta \in \bm{\Delta}$.
We use the tools developed in \cite{hecker2005enhanced} to obtain models in LFT form.
The resulting dynamical system is described by the equations
\begin{align}\label{eq:P_LFT}
   \begin{bmatrix}
   \dot{x}(t) \\
   v(t) \\
   y(t)
   \end{bmatrix}
   & =
   \begin{bmatrix}
      A   & B_w     & B_u     \\
      C_v & D_{vw}  & D_{vu}  \\
      C_y & D_{yw}  & D_{yu}
   \end{bmatrix}
   \begin{bmatrix}
   x(t) \\
   w(t) \\
   u(t)
   \end{bmatrix}, \\
   w(t) & = \Delta~v(t).
\end{align}
\section{Model Reduction via Fixed Structure Synthesis}\label{sec:mode_reduction_tech}
\begin{figure}[]
\centering
    \begin{minipage}{0.23\textwidth}
        	\includegraphics[scale=0.35]{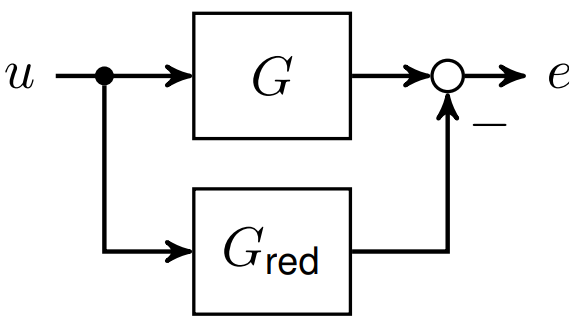}
    \end{minipage}
    \begin{minipage}{0.23\textwidth}
        	\includegraphics[scale=0.35]{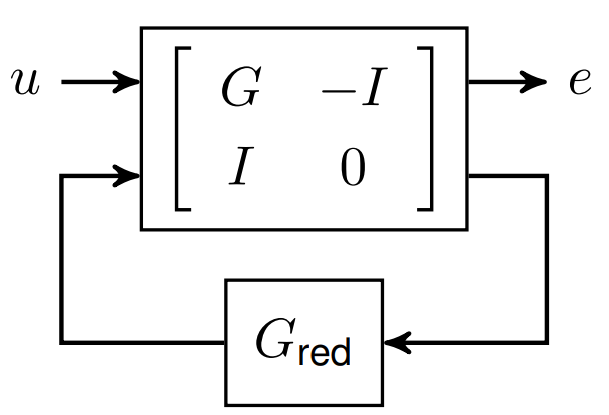}
    \end{minipage}
	\caption{Equivalent block diagrams representing the model approximation problem (left) and a fixed structure controller synthesis problem (right)}
	\label{fig:block_diagram}
\end{figure}

In this section, we present our approach to model reduction in detail.
Standard model reduction techniques (see \cite{Wood.1995}) based on balanced truncation typically involve the following steps.
\begin{enumerate}
    \item First, the following optimization problem is solved to obtain the generalized Gramians $\mathcal{O}_{\rho}$ and $\mathcal{C}_{\rho}$:
\begin{align*}
	\min_{\mathcal{C}_{\rho},\,\mathcal{O}_{\rho}} \hspace{0.5cm} & \textnormal{Trace}\left(\mathcal{C}_{\rho}\mathcal{O}_{\rho}\right) \\
	\textnormal{ subject to }&A^{T}_\rho \mathcal{O}_{\rho}+\mathcal{O}_{\rho} A_\rho+C^{T}_\rho C_\rho<0 \quad \forall \rho \in \mathcal{P}, \\
	&A_\rho \mathcal{C}_{\rho}+\mathcal{C}_{\rho} A^{T}_\rho+B_\rho B^{T}_\rho<0 \quad ~\forall \rho \in \mathcal{P}, 
\end{align*}
where Trace denotes the trace of a matrix.
    \item Next, a transformation matrix $T$ is computed such that
\begin{align*}
		T^T \mathcal{O}_{\rho} T = T^{-1} \mathcal{C}_{\rho} T^{-T} = \Sigma = \begin{bmatrix}
		   \sigma_1,&&\\
		   &\ddots&\\
		   &&\sigma_n
		\end{bmatrix}.
\end{align*}
\item Finally, matrix $T$ is truncated to obtain the reduced model and the model approximation error is available in the form of the sum of the truncated singular values.
\end{enumerate}
Heuristic approaches to minimize the sum of the truncated singular values are then applied to obtain the final reduced model.
Following up on ideas from \cite{beck1997model} and \cite{al2016structure} we propose to directly minimize the error bound by using controller synthesis techniques.
We pose the model reduction problem as an optimal controller synthesis problem using a suitable norm. 
Consider the optimization problem with reduced plant-order $n$:

\begin{align}
    \min_{G_{\textnormal{red}} \in \: \mathcal{G}_n} \hspace{0.5cm} & \left \Vert G-G_{\textnormal{red}}\right \Vert,
\end{align}
where \textcolor{black}{$\mathcal{G}_n$ is the set of LPV systems of order $n$.
}
This is depicted in the block diagram in Fig. \ref{fig:block_diagram} (left).
\textcolor{black}{
Observe that the block diagram in Fig. \ref{fig:block_diagram} (left) can be equivalently transformed to the block diagram in Fig. \ref{fig:block_diagram} (right), i.e., 
\begin{align*}
    G-G_{\textnormal{red}}=\mathcal{F}_l \left(\begin{bmatrix}
   G & -I \\ I & 0
\end{bmatrix},G_{\textnormal{red}}\right).
\end{align*}
We can thus convert the model approximation (model reduction) problem into a controller synthesis problem by defining an appropriate generalized plant
\begin{align*}
G_{\textnormal{gp}}=\begin{bmatrix}
   G & -I \\ I & 0
\end{bmatrix}    
\end{align*}
and posing the controller synthesis problem as
\begin{align}
    \min_{K} \hspace{0.5cm} & \Vert \mathcal{F}_l (G_{\textnormal{gp}},K)\Vert.
\end{align}
Optimal solution $K^*$ to this problem thus produces an approximate model $G_{\textnormal{red}}=K^*$ with an approximation error of $\Vert G-G_{\textnormal{red}} \Vert =\Vert \mathcal{F}_l (G_{\textnormal{gp}},K^*)\Vert$.}
Standard computationally efficient techniques based on linear matrix inequalities (LMI) (see \cite{scherer1999lecture}) provide a solution to the controller synthesis problem when there is no restriction on the order of the controller.
However, this would lead to a $G_{\textnormal{red}}$ which has the same order as that of $G$ and thus not solve the model-reduction problem.
Therefore, we employ fixed-structure synthesis algorithms to solve the above problem that allow us to specify a fixed order of the sought controller $G_{\textnormal{red}}$ which is here the reduced model.
Starting points for this are tools developed in \cite{chughtai2011hybrid,chughtai2010simply,chughtai2007fixed,abbas2008hybrid}.
This avoids computing the tranformation matrices.
Furthermore, since the approximation error is the objective function of the optimization problem, we obtain the error estimates as a direct outcome of the optimization.
The equivalent controller synthesis problem obtained in the previous section leads to a bi-linear matrix inequality (BMI) and cannot be solved by standard LMI solvers.
We use a gradient based optimization algorithm in \cite{abbas2008hybrid} to solve the BMI problem.
The interested reader is pointed to \cite{abbas2008hybrid} for details on the synthesis algorithm.

Finally, owing to the flexibility of the gradient-based fixed-structure synthesis algorithm, we can easily enforce a desired structure in the model matrices of the reduced plant.
This can be done by imposing a suitable structure on the model matrices of the sought reduced model.
\textcolor{black}{Letting $\mathcal{A}\subset \mathbb{R}^{n\times n}$, $\mathcal{B}\subset \mathbb{R}^{n\times n_u}$, $\mathcal{C}\subset \mathbb{R}^{n_y\times n}$ and $\mathcal{D}\subset \mathbb{R}^{n_y\times n_u}$ denote the sets of model matrices with an appropriate size and a specific sparsity structure, we can pose the optimization problem
\begin{align}
    \min_{G_{\textnormal{red}} \in \: \mathcal{G}^s_n} \hspace{0.5cm} & \left \Vert G-G_{\textnormal{red}}\right \Vert,
\end{align}
where
\begin{align*}
    \mathcal{G}^s_n=\left\{ \left[\begin{array}{c|c}
	A_{\rho}     &  B_{\rho}\\
	\hline
	C_{\rho}     &  D_{\rho}
\end{array}\right] | A_{\rho}\in \mathcal{A},B_{\rho}\in \mathcal{B},C_{\rho}\in \mathcal{C},D_{\rho}\in \mathcal{D}\right\}.
\end{align*}}
As briefly discussed earlier, it might be desirable to obtain the state matrix of the reduced plant in a modal form. 
This can easily be enforced in the proposed approach by setting
$\mathcal{A}\subset \mathbb{R}^{n\times n}$ to be the set of block-diagonal matrices with block-size at most 2 and setting $\mathcal{B}=\mathbb{R}^{n\times n_u}$, $\mathcal{C}=\mathbb{R}^{n_y\times n}$ and $\mathcal{D}= \mathbb{R}^{n_y\times n_u}$.
For the considered example in the next section, we obtain two reduced models, viz.,
one without imposing any structure and one where we impose a block-diagonal structure on the state-matrix.
\section{Numerical Results} \label{se:numericalResults}
\begin{figure}[]
	\begin{center}
      \def\svgwidth{\columnwidth}
      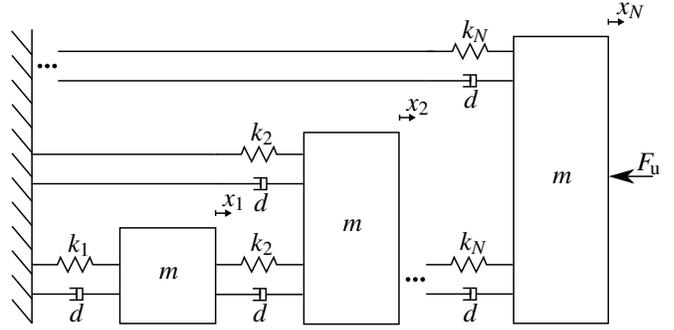   
   \end{center}
	\caption{Mass-spring-damper system containing $N$ blocks with horizontal displacement $x_i$ for $i=1,...,N$ and equal mass ($m = 1\,\text{kg}$), dampers ($d = 0.75\,\text{Ns/m}$) and externally scheduled springs ($k_i = k_0 + \rho_i k_\rho$ with $k_0 = 0.5\,\text{N/m}$, $k_\rho = 0.3\,\text{N/m}$ and $\rho_i \in \mathcal{P} = \left[ -1, 1 \right]$ for $i = 1,...,N$). System input is the external force $F_\text{u}$. System output is the displacement $x_N$.}
   \label{fig:massSpringDamper}
\end{figure}

We illustrate the applicability of the proposed method on an example of chained multiple mass-spring-damper system with time-varying, scheduled stiffness parameters of individual springs, borrowed from the literature~\cite{Theis.2018} which is slightly modified to add more scheduling parameters (see Fig.\,\ref{fig:massSpringDamper}).
The equations of motion for the $N$ blocks are given by
\begin{subequations}
   \begin{align}
      m \ddot{x}_i &= \left\{ \begin{array}{@{}ll@{}}
         -F_1 - F_{1,2}, & \textnormal{ if }i = 1,\\
         -F_i - F_{i,i-1}-F_{i,i+1}, \hspace{.42cm}\rule{0pt}{0pt} & \textnormal{ if } i=2,...,N-1,\\
         -F_N - F_{N,N-1} + F_\text{u}, &  \textnormal{ if }i = N,
      \end{array}\right.\\[3pt]
      F_i &= d\dot{x}_i + k_{i}(\rho_i) x_i,\\
      F_{i,j} &= d(\dot{x}_i - \dot{x}_j) + k_{j}(\rho_j)(x_i-x_j),\\
      y&=x_N.
   \end{align}
\end{subequations}
where the forces $F_i$ and $F_{i,j}$ depend on the externally scheduled spring constants, $k_i(\rho_i)$ for $i = 1,...,N$.
This system can be easily extended in terms of the number of states by  adding more blocks.
Also the number of scheduling parameters can be varied by assuming that either the stiffnesses of all springs are equal ($n_\rho=1$), or that $\rho_i=\rho_{i+n_\rho}$ for $i = 1,...,N-n_\rho$.
For the numerical results, we let $N=10$ to obtain a full-order model (called the original model) of order 20.
The admissible parameter range is taken as ${\cal P} = [-1,1]$.
\begin{table}[h!]
    \caption{\textcolor{black}{$\mathcal{L}_2$ approximation error bounds for reduced models with and without imposed modal structure}}
    \centering
    \begin{tabular}{ |c|c| } 
         \hline
          &  Approximation error bound\\ 
          \hline
         $||G-G_{\textnormal{red}}||$ & 0.3056 \\ 
         $||G-G_{\textnormal{red-modal}}||$ &7.6012 \\ 
         \hline
    \end{tabular}
    \label{table:table_norms}
\end{table}
We apply the proposed model reduction techniques described in Section \ref{sec:mode_reduction_tech} to reduce the original model of order 20 (denoted by $G$) to obtain two reduced-order models of order 4, one without any imposed structure on the state-matrix which we denote by $G_{\textnormal{red}}$, and one with an imposed block-diagonal/modal structure (with block-size $2$) on the state-matrix which we denote by $G_{\textnormal{red-modal}}$.
Table \ref{table:table_norms} shows a comparison of the obtained approximation error bounds. 
We can see that imposing structure on the state-matrix comes at the cost of a larger approximation error.
These reduced-order models are next compared with the original model in time and frequency domain in the next sub-section.
\subsection{Comparison of open-loop models}
\begin{figure}[]
	\centering
    \input{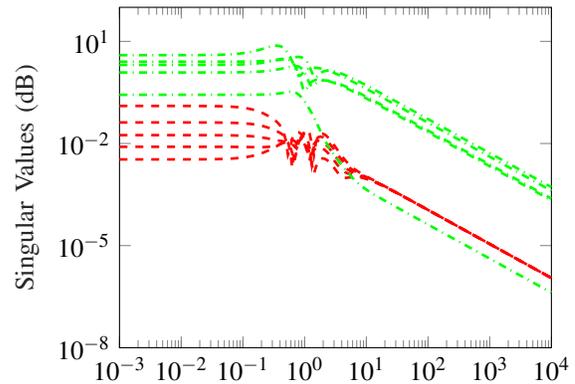}
	\caption{Frequency response of the error system $G-G_{\textnormal{red}}$ (shown with dashed red curves) and the error system $G-G_{\textnormal{red-modal}}$ (shown with dash-dotted green curves). The different curves correspond to the different values of the scheduling parameter \textcolor{black}{$\rho$ from the grid $\{-1,-0.5,0,0.5,1\}$}.}
	\label{fig:freq_response_error}
\end{figure}

\begin{figure}[]
	\centering
    \input{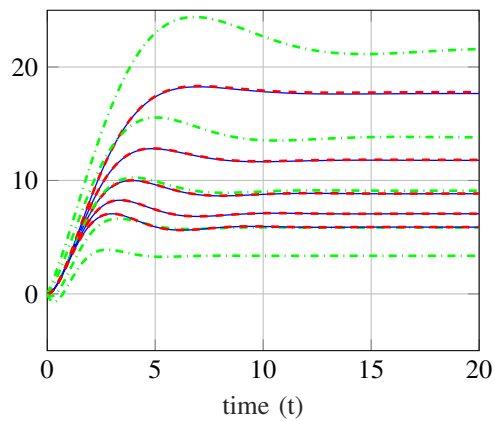}
	\caption{Step response of the original model $G$(shown with solid blue curves), reduced model $G_{\textnormal{red}}$ (shown with dashed red curves) and the reduced model $G_{\textnormal{red-modal}}$ (shown with dash-dotted green curves). The different curves correspond to the different values of the scheduling parameter \textcolor{black}{$\rho$ from the grid $\{-1,-0.5,0,0.5,1\}$}.}
	\label{fig:step_response}
\end{figure}

We first compare the different models in frequency domain by looking at the sigma plots in Fig. \ref{fig:freq_response}.
Multiple curves of the same color show the sigma plots of the same LPV model evaluated at 5 uniformly spaced grid points on ${\cal P} = [-1,1]$. 
It can be seen that the reduced model without any imposed structure $G_{\textnormal{red}}$ (shown in red) matches well with the original model $G$ (blue curves).
Although, the reduced model $G_{\textnormal{red-modal}}$ with an imposed block-diagonal modal structure on the state-matrix does not match well with original model $G$, the inaccuracy is dominant in the poor approximation of the static-gain.
To bring out the comparison clearly, Fig. \ref{fig:freq_response_error} shows the sigma plots of the error system ($G-G_{\textnormal{red}}$) (shown in red) and the error system ($G-G_{\textnormal{red-modal}}$) (shown in green).
It can be observed that the imposing the modal structure in the state-matrix costs us an inaccurate model bringing out the inherent trade-off.
Finally, Fig. \ref{fig:step_response} shows that the step response of $G_{\textnormal{red}}$ matches well with the original model.
At the same time, the step response of $G_{\textnormal{red-modal}}$ does not match well with the original model in terms of static gain.

It is apparent in the above plots that the mismatch between the original model and the reduced model $G_{\textnormal{red-modal}}$ is concentrated in the low-frequency (static gain) regime.
It is known that a model used for controller design needs to be accurate around the bandwidth and an uncertain static-gain of the open loop can be handled by a well-designed controller.
This is evident in the comparison of closed-loop performance discussed in the next sub-section.
\begin{figure}[]
	\centering
    \input{figures/sigma_plots_open_loop}
	\caption{Frequency response of the original model $G$ (shown with solid blue curves), reduced model $G_{\textnormal{red}}$) (shown with dashed red curves) and the reduced model with a block diagonal structure in state-matrix $G_{\textnormal{red-modal}}$ (shown with dash-dotted green curves). The different curves correspond to the different values of the scheduling parameter \textcolor{black}{$\rho$ from the grid $\{-1,-0.5,0,0.5,1\}$}.}
	\label{fig:freq_response}
\end{figure}
\subsection{Comparison in closed-loop}
In this section, we use standard LPV synthesis techniques to design a controller for the reduced model and test it on the original model to compare performance in closed-loop.
We give only a brief summary of the main ideas behind the synthesis techniques.
It is assumed that an LPV model of the corresponding plant is available in LFT form as in \eqref{eq:P_LFT}
where $u$ is the control input and $y$ is the measured signal. 
Standard LPV synthesis techniques are used to design controllers for the original full-order plant and the reduced plants and we refer the interested reader to \cite{scherer1997full} for details. 
The LPV controller has the form described by
\begin{align} \label{eq:K_LFT}
   \begin{bmatrix}
   \dot{x}_{\text{\tiny K}}(t) \\
   v_{\text{\tiny K}}(t) \\
   u(t)
   \end{bmatrix}
   & =
   \begin{bmatrix}
      A^{\text{\tiny K}}        & B^{\text{\tiny K}}_{w}    & B^{\text{\tiny K}}_{y}    \\
      C^{\text{\tiny K}}_{v}    & D^{\text{\tiny K}}_{vw}   & D^{\text{\tiny K}}_{ve}   \\
      C^{\text{\tiny K}}_{u}    & D^{\text{\tiny K}}_{uw}   & D^{\text{\tiny K}}_{ue} 
   \end{bmatrix}
   \begin{bmatrix}
   x_{\text{\tiny K}}(t) \\
   w_{\text{\tiny K}}(t) \\
   e(t)
   \end{bmatrix} \\
   w_{\text{\tiny K}}(t) & = \Delta_{\text{\tiny K}}~v_{\text{\tiny K}}(t).
\end{align}
where $e=r-y$ with the external signal $r$ as the reference command.
This controller is interconnected with the plant through the signals $u$ and $e$ along with a performance channel $z$ (typically incorporating $S/KS$ loop shaping) which results in an LPV closed-loop model in LFT form described by
\begin{align} \label{eq:CL_LFT}
   \begin{bmatrix}
   \dot{x}_{\text{\tiny CL}} \\
   v_{\text{\tiny CL}} \\
   z
   \end{bmatrix}
   & =
   \begin{bmatrix}
      \mathcal{A}   & \mathcal{B}_w     & \mathcal{B}_r     \\
      \mathcal{C}_v & \mathcal{D}_{vw}  & \mathcal{D}_{vr}  \\
      \mathcal{C}_z & \mathcal{D}_{zw}  & \mathcal{D}_{yr}
   \end{bmatrix}
   \begin{bmatrix}
   x_{\text{\tiny CL}} \\
   w_{\text{\tiny CL}} \\
   r
   \end{bmatrix} \\
   w_{\text{\tiny CL}} & = \Delta_{\text{\tiny CL}}~v_{\text{\tiny CL}},
\end{align}
where $x_{\text{\tiny CL}}=\left[x,\,x_{\text{\tiny K}}\right]^\text{\tiny T}$,  $\Delta_{\text{\tiny CL}}$ collects in block diagonal fashion the plant and controller block matrices, $\Delta$ and $\Delta_{\text{\tiny K}}$ respectively and the calligraphic system matrices contain the resulting closed-loop interconnection of the LPV plant and controller. 

Feasibility of a set of parameterized LMIs (representing the LPV version of the bounded real lemma, see \cite{scherer1997full} for details) guarantees stability and induced $\mathcal{L}_2$ performance, via the existence of the corresponding Lyapunov matrix and multipliers.
To reduce the complexity of the controller synthesis, a parameter-independent Lyapunov function is used along with D/G multipliers. 
The latter guarantees the parameter block of the controller be an exact copy of the one of the plant.
We apply the Full-Block S-Procedure (FBSP) and the parameter elimination lemma \cite{scherer1997full} in the form of a two-step controller synthesis procedure to obtain the controller.
In the first step, an LMI optimization problem is solved to obtain Lyapunov matrices and structured multiplier matrices, which guarantee a level of performance. 
In the second step, the state-space matrices of the LPV controller can be constructed using the obtained matrices from the first step.

\begin{figure}[]
	\centering
    \input{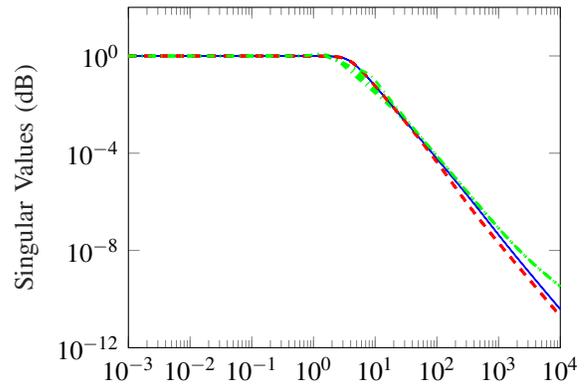}
	\caption{Closed-loop frequency response of the original model (shown with solid blue curves), reduced model (shown with dashed red curves) and the reduced model with a block diagonal structure in $A$ (shown with dash-dotted green curves). The different curves correspond to the different values of the scheduling parameter \textcolor{black}{$\rho$ from the grid $\{-1,-0.5,0,0.5,1\}$}.}
	\label{fig:cl_freq_response}
\end{figure}

\begin{figure}[]
	\centering
    \input{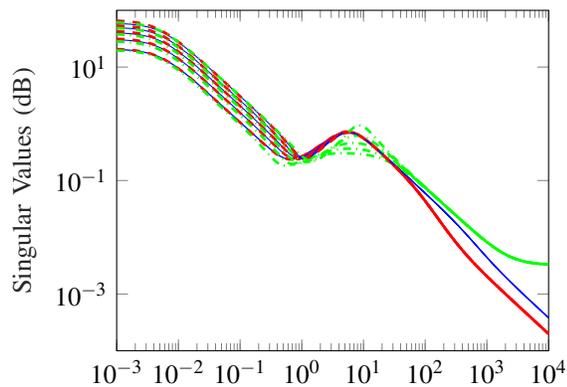}
	\caption{Sigma plots of the controller for original model (shown with solid blue curves), reduced model (shown with dashed red curves) and the reduced model with a block diagonal structure in $A$ (shown with dash-dotted green curves). The different curves correspond to the different values of the scheduling parameter \textcolor{black}{$\rho$ from the grid $\{-1,-0.5,0,0.5,1\}$}.}
	\label{fig:sigma_controller}
\end{figure}

We emphasize that the reduced model is used solely for controller design and the obtained controller is then tested in closed-loop with the original plant.
Let $K$, $K_{\textnormal{red}}$ and $K_{\textnormal{red-modal}}$ be the controllers designed using the plants $G$, $G_{\textnormal{red}}$ and $G_{\textnormal{red-modal}}$, respectively.
These controllers are implemented in closed-loop with the original plant $G$ and compared in the following to evaluate the effect of model reduction on closed-loop performance.
We start the assessment of the closed-loop by looking at the sigma plots of the closed-loop in Fig. \ref{fig:cl_freq_response}.
Fig. \ref{fig:cl_freq_response} shows the sigma plots of the closed-loop system with controllers $K$, $K_{\textnormal{red}}$ and $K_{\textnormal{red-modal}}$.
Since all curves are concentrated at $1$ at low frequencies, we can conclude that the closed-loop performance is good in this frequency regime with both reduced models.
This also agrees with the step-response curves shown in Fig. \ref{fig:cl_step_response} where all closed-loop models show a zero steady-state error.
Furthermore, this illuminates the fact that in spite of the poor open-loop approximation error at low frequencies, the designed controller is able to handle this and achieve a good performance in closed-loop at low-frequencies.
The transient responses of the closed-loop with a controllers $K$ and $K_{\textnormal{red}}$ match well and show good performance whereas the transient response of the closed loop with controller $K_{\textnormal{red-modal}}$ designed using the reduced model $G_{\textnormal{red-modal}}$ shows a  performance degradation.
Specifically, we get a slightly higher rise-time, a higher overshoot and a higher settling time.
Finally, Fig. \ref{fig:sigma_controller} shows the sigma plots of the obtained controllers.
It can be seen that the controllers designed with the different models show very similar frequency responses in general.
While the controllers $K$ and $K_{\textnormal{red}}$ are strictly proper, the controller $K_{\textnormal{red-modal}}$ designed with the reduced order model $G_{\textnormal{red-modal}}$ is bi-proper.
\begin{figure}[]
	\centering
    \input{figures/cl_loop_step_response}
	\caption{Closed-loop step response of the original model (shown with solid blue curves), reduced model (shown with dashed red curves) and the reduced model with a block diagonal structure in $A$ (shown with dash-dotted green curves). The different curves correspond to the different values of the scheduling parameter \textcolor{black}{$\rho$ from the grid $\{-1,-0.5,0,0.5,1\}$}.}
	\label{fig:cl_step_response}
\end{figure}

\section{Conclusions and Future Work} \label{se:conclusion}
We propose a novel model reduction technique for LPV systems by leveraging available tools on fixed-structure synthesis. 
Owing to the flexibility of the used fixed-structure synthesis tools, we are able to impose a desired structure on the model matrices of the reduced-order model. 
We demonstrate the applicability of the results on a benchmark example by first analyzing the performance in open-loop.
Finally, we evaluate the effect of the model reduction technique on controller design by evaluating the closed-loop performance.


 \addtolength{\textheight}{-4cm}   
                                   sure that you do not shorten the textheight too much.

\printbibliography

@book{beck1997model,
  title={Model reduction and minimality for uncertain systems},
  author={Beck, Carolyn L},
  year={1997},
  publisher={California Institute of Technology}
}

@article{al2016structure,
  title={Structure-preserving model reduction for spatially interconnected systems with experimental validation on an actuated beam},
  author={Al-Taie, Fatimah and Werner, Herbert},
  journal={International Journal of Control},
  volume={89},
  number={6},
  pages={1248--1268},
  year={2016},
  publisher={Taylor \& Francis}
}

@article{abbas2008hybrid,
  title={A hybrid gradient-LMI algorithm for solving BMIs in control design problems},
  author={Abbas, Hossam and Chughtai, Saulat S and Werner, Herbert},
  journal={IFAC Proceedings Volumes},
  volume={41},
  number={2},
  pages={14319--14323},
  year={2008},
  publisher={Elsevier}
}

@article{chughtai2010simply,
  title={Simply structured controllers for parameter varying distributed systems},
  author={Chughtai, Saulat S and Werner, Herbert},
  journal={Smart materials and structures},
  volume={20},
  number={1},
  pages={015006},
  year={2010},
  publisher={IOP Publishing}
}

@article{chughtai2007fixed,
  title={Fixed structure controller design for a class of spatially interconnected systems},
  author={Chughtai, Saulat S and Werner, Herbert},
  journal={IFAC Proceedings Volumes},
  volume={40},
  number={9},
  pages={224--229},
  year={2007},
  publisher={Elsevier}
}

@article{chughtai2011hybrid,
  title={A hybrid approach to the synthesis of simply structured robust and gain-scheduled controllers},
  author={Chughtai, Saulat S and Werner, Herbert},
  journal={Applied Soft Computing},
  volume={11},
  number={6},
  pages={4078--4086},
  year={2011},
  publisher={Elsevier}
}

@article{hoffmann2014survey,
  title={A survey of linear parameter-varying control applications validated by experiments or high-fidelity simulations},
  author={Hoffmann, Christian and Werner, Herbert},
  journal={IEEE Transactions on Control Systems Technology},
  volume={23},
  number={2},
  pages={416--433},
  year={2014},
  publisher={IEEE}
}

@article{hecker2005enhanced,
  title={Enhanced LFR-toolbox for MATLAB},
  author={Hecker, Simon and Varga, Andras and Magni, Jean-Fran{\c{c}}ois},
  journal={Aerospace Science and Technology},
  volume={9},
  number={2},
  pages={173--180},
  year={2005},
  publisher={Elsevier}
}

@inproceedings{theis2015modal,
  title={Modal matching for LPV model reduction of aeroservoelastic vehicles},
  author={Theis, Julian and Takarics, B{\'e}la and Pfifer, Harald and Balas, Gary J and Werner, Herbert},
  booktitle={AIAA Atmospheric Flight Mechanics Conference},
  pages={1686},
  year={2015}
}

@inproceedings{heeren2022grid,
  title={Grid-Free Constraints for Parameter-Dependent Generalized Gramians via Full Block S-Procedure},
  author={Heeren, Lennart and Werner, Herbert},
  booktitle={2022 European Control Conference (ECC)},
  pages={1073--1078},
  year={2022},
  organization={IEEE}
}

@article{amsallem2011online,
  title={An online method for interpolating linear parametric reduced-order models},
  author={Amsallem, David and Farhat, Charbel},
  journal={SIAM Journal on Scientific Computing},
  volume={33},
  number={5},
  pages={2169--2198},
  year={2011},
  publisher={SIAM}
}

@article{shamma2012overview,
  title={An overview of LPV systems},
  author={Shamma, Jeff S},
  journal={Control of linear parameter varying systems with applications},
  pages={3--26},
  year={2012},
  publisher={Springer}
}

@inproceedings{scherer1997full,
  title={A full block S-procedure with applications},
  author={Scherer, Carsten W},
  booktitle={Proceedings of the 36th IEEE Conference on Decision and Control},
  volume={3},
  pages={2602--2607},
  year={1997},
  organization={IEEE}
}

@article{scherer1999lecture,
  title={Lecture notes DISC course on linear matrix inequalities in control},
  author={Scherer, Carsten and Weiland, Siep},
  journal={Delft University},
  year={1999},
  publisher={Citeseer}
}

@article{lall2003structure,
  title={Structure-preserving model reduction for mechanical systems},
  author={Lall, Sanjay and Krysl, Petr and Marsden, Jerrold E},
  journal={Physica D: Nonlinear Phenomena},
  volume={184},
  number={1-4},
  pages={304--318},
  year={2003},
  publisher={Elsevier}
}

@article{Theis.2018,
 author = {Theis, Julian and Seiler, Peter and Werner, Herbert},
 year = {2018},
 title = {{LPV model order reduction by parameter-varying oblique projection}},
 pages = {773--784},
 volume = {26},
 number = {3},
 journal = {{IEEE Trans. Contr. Syst. Technol.}}
}

@phdthesis{Wood.1995,
 author = {Wood, G. D.},
 year = {1995},
 title = {{Control of parameter-dependent mechanical systems}},
 address = {Cambridge},
 school = {{St Johns College}},
 type = {{Dissertation}}
}

@book{Zhou.1996,
 author = {Zhou, Kemin and Doyle, John Comstock and Glover, Keith},
 year = {1996},
 title = {{Robust and optimal control}},
 address = {Upper Saddle River, NJ},
 publisher = {{Prentice Hall}}
}

\end{document}